\documentclass[12pt,prc,aps,superscriptaddress]{revtex4}   
\usepackage{graphicx}
\tighten
\begin{document}
\draft
\title{                                                     
Microscopic in-medium nucleon-nucleon cross sections                       
with improved Pauli blocking effects}              
\author{            
B. Chen}                                                                                
\affiliation{                 
Physics Department, University of Idaho, Moscow, ID 83844-0903, U.S.A  } 
\author{            
F. Sammarruca}                                                                                
\affiliation{                 
Physics Department, University of Idaho, Moscow, ID 83844-0903, U.S.A  } 
\author{C.A. Bertulani} 
\affiliation{                 
Department of Physics and Astronomy, 
Texas A\&M University-Commerce, Commerce, TX 75428 } 
\begin{abstract}
We present updated predictions of effective elastic nucleon-nucleon cross sections 
intended for use in nucleus-nucleus reactions.                
A novel characteristic of the present approach combines all microscopic medium effects included in the Dirac-Brueckner-Hartree-Fock 
G-matrix with a Pauli blocking mechanism which is more appropriate for 
applications in ion-ion reaction models as compared to a previous approach.      
The effective in-medium cross section is found to be quite sensitive to the 
description of Pauli blocking in the final configurations. 

\end{abstract}
\pacs {21.65.+f, 21.30.Fe} 
\maketitle

\section{Introduction} 

The investigation of the effective nucleon-nucleon (NN) interaction in  dense 
hadronic matter is a topic of fundamental importance for nuclear reactions at intermediate energies ($20 \ {\rm MeV/nucleon} \lesssim E_{lab} \lesssim 300 \ {\rm MeV/nucleon}$) and for nuclear structure in general.                    
The relevant literature is very vast. Reference~\cite{MP} is just a representative example of the traditional microscopic
approach where two-nucleon correlations in nuclear systems are introduced through the G-matrix.                                     
Moreover, the effective NN interaction is the main ingredient of microscopic predictions of the nuclear equation 
of state (EoS) and thus impacts the properties of compact stars. Dense hadronic matter 
can also be created in the laboratory 
in energetic heavy-ion (HI) collisions. Simulations of HI collisions are typically 
based on transport equations and describe the evolution of a non-equilibrium system of strongly
interacting hadrons undergoing two-body collisions in the presence of a mean field.                    
The Boltzmann-Uehling-Uhlenbeck equation \cite{BUU1,BUU2} and quantum molecular dynamics \cite{QMD}, 
along with their relativistic counterparts \cite{rel1,rel2,rel3}, have been typically employed to describe 
intermediate-energy HI reactions. 
{\it In-medium} two-body cross sections are therefore an important component of such simulations. 

In direct reactions at intermediate energies the NN cross sections are often  used as input to obtain quantum refractive and diffractive effects, replacing the role of optical potentials commonly used in low energy reactions \cite{BD04}. Examples such as knockout (stripping and diffraction dissociation) reactions, elastic scattering, charge-exchange, and excitation of giant resonances, are often carried out using reaction mechanisms based on the construction of scattering matrices built from the underlying NN scattering. Reaction calculations at intermediate to high energy are often conducted within the framework
of the Glauber approximation \cite{Glauber} and have been a frequent tool       
for testing nuclear models and constraining nuclear sizes. 
In fact, the description of complex nuclear reactions at intermediate energies 
based on individual NN collisions has a long tradition. 
In the framework of the Glauber model, the reaction cross section is written in terms 
of the ``thickness function", which 
is the product of the averaged NN cross section and the overlap
integral of the target and projectile local densities.                                         

In-medium NN cross sections have been calculated with a variety of methods. 
In semi-phenomenological approaches, one makes the assumption that the transition matrix in the medium is approximately 
the same as the one in vacuum and that medium effects come in only through the use of effective masses
in the phase space factor \cite{Pand,Pers,Li05}. Then, the in-medium cross section is scaled (relative to its 
value in vacuum) as the square of the ratio of the (reduced) masses.          
Phenomenological formulas, such as the one in Ref.~\cite{sigmed}, have been developed for practical purposes 
and combine the energy dependence 
of empirical free-space NN cross sections with the density dependence of some microscopic models. 

Microscopic predictions based on a medium-modified collision matrix were reported, for instance, in Ref.~\cite{LM},
where Dirac-Brueckner-Hartree-Fock (DBHF) medium effects were applied to obtain a medium-modified $K$-matrix. 
More recent microscopic calculations applied DBHF medium effects to produce a                     
complex G-matrix including consideration of isospin dependence 
in asymmetric nuclear matter \cite{SK}. 

It is the purpose of this paper to present our updated predictions of microscopic in-medium elastic NN cross sections    
with an improved description of Pauli blocking. The main objective is to produce two-body cross sections
which include, microscopically, all important medium effects and are suitable for realistic applications in  
nucleus-nucleus scattering at intermediate energies including direct and central collisions. 
As explained in Section II, we start from a one-boson-exchange NN potential, 
which describes well the elastic part of the NN interaction up to high energy.   
Thus, as long as we are not interested in pion production, which is negligible up
to, at least, several hundreds of MeV, it is reasonable to use NN elastic cross sections 
as input to the reaction model. Of course, the elastic part of the 
NN interaction can and does generate inelastic nucleus-nucleus scattering. 

In Section II, we describe the details of the calculation and highlight the differences 
with our previous approach. We then present a selection of results (Section III) followed by our conclusions and
outlook (Section IV).

\section{Description of the calculation} 
\subsection{The Dirac-Brueckner-Hartree-Fock G-matrix} 

The starting point of our calculation is a realistic NN interaction which is applied in the 
nuclear medium without any additional free parameters. 
We use                                
relativistic meson theory, which we find to be an appropriate framework to deal with the high momenta encountered in 
dense matter. In particular, 
the one-boson-exchange (OBE) model has proven very successful in describing NN data in free space 
and has a good theoretical foundation. 

The OBE potential is defined as a sum of one-particle-exchange amplitudes of certain bosons with 
given mass and coupling. In general, six non-strange bosons with masses below 1 GeV$/c^2$ are used.
Thus,
\begin{equation}
V = \sum_{\alpha=\pi,\eta,\rho,\omega,\delta,\sigma} V_{\alpha}^{OBE}  \;,
\label{OBE}
\end{equation}
with $\pi$ and $\eta$ pseudoscalar, $\sigma$ and $\delta$ scalar, and $\rho$ and $\omega$ vector
particles. For more details, see Ref.~\cite{Mac89}.    Among the many available OBE potentials, some being part of the ``high-precision generation" \cite{pot1,pot2}, 
we seek a momentum-space potential developed within a relativistic scattering equation, such as the 
one obtained through the Thompson \cite{Thom} three-dimensional reduction of the Bethe-Salpeter equation \cite{BS}. 

First, a self-consistent calculation of (symmetric or asymmetric) nuclear matter is performed 
within the DBHF approach \cite{FS10}. This step yields, along with the EoS, the self-consistent nuclear matter
potential, which is conveniently parametrized in terms of nucleon effective masses (see Ref.~\cite{FS10}
for details). Then, the 
Thompson equation is solved for two nucleons scattering at some positive energy in the presence of a mean 
field due to the medium. The presence of the medium is accounted for through the (previously calculated) effective masses
(applied in the two-nucleon propagator and also in the Dirac spinors representing the nucleons, consistent with the DBHF philosophy) and 
the presence of the (angle-averaged) Pauli operator
to account for Pauli blocking of the {\it intermediate states}.  

In the usual free-space scattering scenario, the two-body cross section is typically represented as a function 
of the incident laboratory energy, uniquely related to the nucleon momentum in the two-body
center-of-mass frame, ${\bf q}$, through relativistic invariants which yield the well-known relation $E_{lab} = 2q^2/m$. 
In nuclear matter, though, the Pauli operator depends also on the total momentum of the two nucleons
in the nuclear matter rest frame. 
For simplicity, in the past we have used in-vacuum kinematics to define the total momentum of the 
two-nucleon system (that is, we assumed that the target nucleon is at rest, on the average).
Schematically, the effect of Pauli principle on intermediate states arises in the G-matrix through the in-medium scattering equation \cite{GWW58}:
\begin{equation}
\langle\mathbf{k}|\mathrm{G}(\mathbf{p})|\mathbf{q}_{0}\rangle
=\langle\mathbf{q}|\mathrm{V}|\mathbf{q}_{0}\rangle
-\int{\frac
{d^{3}q^{\prime}}{(2\pi)^{3}}}{\frac{\langle\mathbf{q}|\mathrm{V}%
|\mathbf{q^{\prime}}\rangle Q(\mathbf{q^{\prime}},\mathbf{p}%
)\langle\mathbf{q^{\prime}}|\mathrm{G}(\mathbf{p})|\mathbf{q}%
_{0}\rangle}{E(\mathbf{p},\mathbf{q^{\prime}})-E_{0}-i\epsilon}} \; , 
\label{10}%
\end{equation}
with $\mathbf{q}_{0}$, $\mathbf{q}$, and $\mathbf{q^{\prime}}$ the
initial, final, and intermediate relative momenta of the NN pair in their center of mass,
 and ${\bf p}$ their total momentum. 
$E$ is the energy of the two-nucleon system in the center-of-mass, and $E_{0}$ is the same quantity on-shell. 

To account for Pauli blocking of the {\it final state}, 
we define the total elastic cross section as 
\begin{equation}
{\bar \sigma}_{NN}(q) = \int \Big (\frac{d \sigma}{d \Omega} \Big)^{DBHF} Q(q,p,\theta,\rho)  \; d \Omega \; , \label{sigmed}
\end{equation} 
where $({d \sigma / d \Omega})^{DBHF}$ is the elastic differential cross section obtained from the G-matrix amplitudes as described above.  
$\theta$ is the 
scattering angle and $k_F$ the Fermi momentum.             
The presence of the Pauli operator in Eq.~(\ref{sigmed}) signifies that the integration domain is restricted by \cite{SK}     
\begin{equation}
\frac{k_F^2 -p^2 -q^2}{2pq} \le cos \; \theta \le                                                               
\frac{p^2 +q^2 - k_F^2}{2pq} \; .                                                                             
\label{cos}
\end{equation} 
Setting $Q=1$ in Eq.~(\ref{sigmed}) amounts to ignoring Pauli blocking of the final state. (The virtual intermediate states
are always subjected to Pauli blocking during the G-matrix calculation which produces the amplitudes 
contained in $(d \sigma / d\Omega)^{DBHF}$.) 
Additional simplifications result from
the assumption that the differential cross section is isotropic. 

\begin{figure}
\centering            
\scalebox{0.46}{\includegraphics{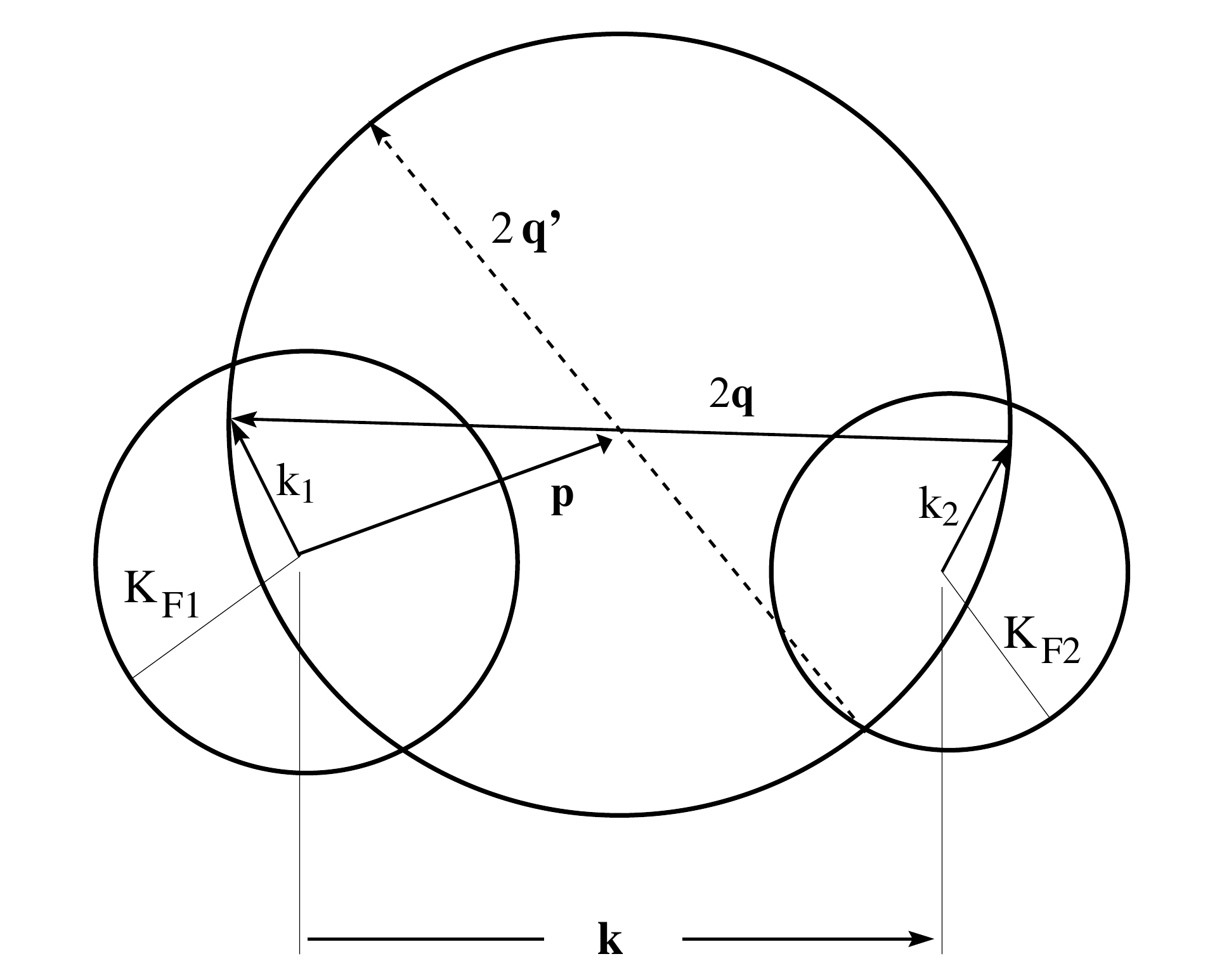}}
\caption{                                                                                                  
Geometrical representation of Pauli blocking.                        
} 
\label{one}
\end{figure}
\subsection{The average in-medium cross section and Pauli blocking effects} 

The cross section defined in Eq.~(\ref{sigmed}) refers to an 
idealized scenario where a projectile nucleon, with some momentum above the bottom of the 
Fermi sea, strikes a target nucleon while both embedded in an infinite medium.
For application to a realistic nucleus-nucleus scattering scenario, it is best to consider
the situation depicted in Fig.~\ref{one}, 
where the two Fermi spheres represent the local densities of the target and projectile ions. 
${\bf k}$ is the incident momentum (the momentum of the colliding {\it nuclei} relative to 
each other), whereas 
${\bf k_1}$ and                                                               
${\bf k_2}$ are the momenta of any two {\it nucleons}. That is,               
${\bf k_1}$ and                                                               
${\bf k_2} + {\bf k}$ are the momenta of the two nucleons with respect to the same point. 
Then, the relative momentum $2{\bf q}$ and the total momentum 
$2{\bf p}$ are given by           
$2{\bf q} = {\bf k}_2 + {\bf k} - {\bf k}_1$, and                          
$2{\bf p} = {\bf k}_1 + {\bf k}_2 + {\bf k}$, respectively.                
The larger circle in the figure is centered at ${\bf p}$   
while                                                                                                       
$|{\bf q}|$ is the radius of the scattering sphere.                       
The vector 
$2{\bf q}$ can rotate around the scattering sphere while maintaining 
constant magnitude due to energy-momentum conservation. 

Notice that, with the definitions given above, relative momenta which are off 
the symmetry axis of the two Fermi spheres (the ${\bf k}$ direction) are allowed, which is
not the case with the assumptions made in Eq.~(\ref{sigmed}). That is, the two interacting nucleons can have
momenta in arbitrary directions. 
In turn, this impacts the solid angle allowed by Pauli blocking, as shown below. 
(For completeness, we provide a detailed derivation of the allowed solid angle in the next subsection.) 

In preparation for the nucleus-nucleus calculation, it is        
shown in Refs.~\cite{HRB,Bert86} that the {\it average} NN cross section (assuming isotropy of the NN differential cross section) can be written as                       
\begin{equation}
{\bar \sigma}_{NN}(k) = \frac{1}{V_{F1}V_{F2}} \int d{\bf k}_1 d{\bf k}_2 \frac{2q}{k}\sigma_{NN}(q) \int _{Pauli} d \Omega \; , 
\label{sigav}
\end{equation} 
where $k_1$ and $k_2$ are smaller $k_{F1}$ and
$k_{F2}$, respectively, and the angular integrations 
extend over all possible directions of 
${\bf k_1}$ and                                                               
${\bf k_2}$ allowed by Pauli blocking. Often, the empirical free-space NN cross section is used in the integral. In our case, $\sigma_{NN}(q)=\sigma^{DBHF}(q)$ is the (microscopic) NN cross section 
which contains additional medium effects
as described in the previous subsection. 
$V_{F1}$ and $V_{F2}$ are the volumes of the two (in general different) Fermi spheres.                                  
Because of azimuthal asymmetry, Eq.~(\ref{sigav}) can be reduced to a fivefold integration.
Notice that the ``symmetric" choice ${\bf q}={\bf p}={\bf k}/2$ \cite{Bert01} amounts to making 
the approximations we adopted when writing Eq.~(\ref{sigmed}).

Finally, for an actual nucleus-nucleus scattering with given $E/A$, the average cross section
given above becomes a function of the laboratory energy, $E(k)$, and 
the local densities of the colliding nuclei, $\rho_i={2 k_{Fi}^3}/({3 \pi^2})$, 
and are ready to be used in typical high-energy calculations. 
This is usually done by defining the average nucleon-nucleon 
cross section at the distance of closest approach ${\bf b}$ between the projectile and the target as      
\begin{equation}
\left\langle \sigma_{NN} (E,b) \right\rangle = \frac{\int d^3 r_1 \ \rho_1({\bf r}_1) \rho_2({\bf r}_1+{\bf b}) \ \sigma_{NN} (E, \rho_1,\rho_2)}{\int d^3 r_1 \rho_1({\bf r}_1) \rho_2({\bf r}_1+{\bf b})} , \label{avesigEb}
\end{equation}
where $\rho_i$ is the local density (at point ${\bf r}$) inside nucleus $i$ and $ \sigma_{NN} (E, \rho_1,\rho_2)$ is the in-medium NN cross section.

The calculation of reaction cross sections in high-energy 
collisions is best described in the eikonal fomalism. The  ``survival amplitudes" (or S-matrices) in 
the eikonal approximation are given by \cite{Glauber,HRB}
\begin{equation}
S_i(E,b)=\exp \left[- \frac{\left<\sigma_{NN}(E,b)\right>}{4\pi}\int_{0}^{\infty}dq\ q\
\rho_{1}\left(  q\right) \rho_{2}\left(  q\right)    J_{0}\left(  qb\right) \right]
\ ,\label{eikphase}%
\end{equation}
where $\rho_{1,2}\left(  q\right)  $ is the Fourier transform of the nuclear
densities of the projectile and target, and the reaction cross sections are 
\begin{equation}
\sigma_R=2\pi \int db \ b \left[ 1- \left| S(b)\right|^2 \right].
\label{sigmaR} 
\end{equation}
Applications to stable and unstable nuclei using Eqs.~(\ref{avesigEb}-\ref{sigmaR}) and our new prescription of Pauli-blocking effects
will be the subject of a future work.

\subsection{Derivation of the Pauli-allowed solid angle}                      
As mentioned in the previous section, the relative momentum $2{\bf q}$ and the total momentum $2{\bf p}$ are given as $2{\bf q} = {\bf k_2}+{\bf k}-{\bf k_1}$, and $2{\bf p} = {\bf k_{1}} + {\bf k_{2}} + {\bf k}$. We also define a vector $2{\bf b}$ as $2{\bf b} = {\bf k_2}+{\bf k_1}-{\bf k}$. Assuming that the collision is elastic, conservation of energy and momentum requires 
\begin{equation}
\begin{array}{rcl}2{\bf p} & = & {\bf k_{1}^{ '}}+{\bf k_{2}^{ '}}+{\bf k} \\ 2{\bf q^{ '}}& = & {\bf k_{2}^{ '}}-{\bf k_{1}^{ '}}+{\bf k} \\ 2{\bf b} & = &{\bf k_{1}^{ '}}+{\bf k_{2}^{ '}}-{\bf k} \; .\end{array}       
\label{eq6}
\end{equation}
The quantities ${\bf k_{1}^{ '}}$ and ${\bf k_{2}^{ '}}$ are the momenta of two nucleons after the collision, whereas ${\bf q^{'}}$ is the relative momentum after collision, with $|{\bf q^{'}}| = |{\bf q}|$. Because of the Pauli exclusion principle, the following restrictions apply: 
\begin{equation}
\begin{array}{rcl}|{\bf k_{1}^{'}}| & = & |{\bf p}-{\bf q^{'}}| > k_{F1} \\ |{\bf k_{2}^{'}}|& = & |{\bf b}+{\bf q^{'}}| > k_{F2} \; ,\end{array}       
\label{eq7}
\end{equation}
or,     
\begin{equation}
\begin{array}{rcl} p^{2}+q^{2}-2pq\cos \alpha_1 & > & k_{F1}^{2} \\ b^{2}+q^{2}+2bq\cos \alpha_2 & > & k_{F2}^2 \; .\end{array}      
\label{eq8}
\end{equation}
\begin{figure}
\centering
\scalebox{0.8}{\includegraphics{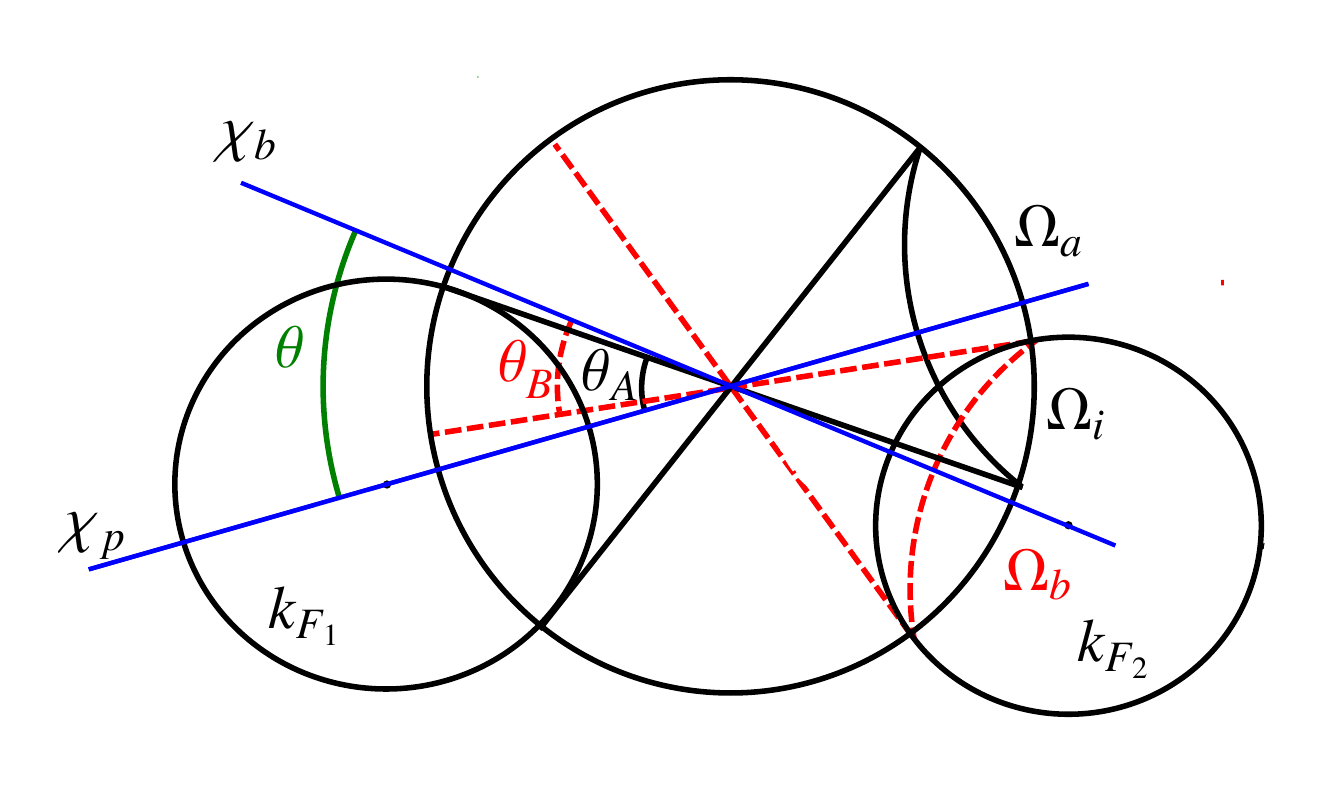}}
\caption{(Color online) Two-dimensional projection of the geometry of Pauli blocking.} 
\label{two}
\end{figure}
In the equations above, 
$\alpha_1$ is the angle between ${\bf p}$ and ${\bf q^{'}}$, and $\alpha_2$ the angle between ${\bf b}$ and ${\bf q^{'}}$. As illustrated in Fig.~\ref{two}, we have     
\begin{equation}
\cos\theta_{A} =  \frac{p^{2}+q^{2}-k_{F1}^{2}}{2pq}, \qquad \cos \theta_{B}  =  \frac{b^{2}+q^{2}-k_{F2}^{2}}{2bq} \; ,
\label{eq9} 
\end{equation}
with $\theta_A$ and $\theta_B$ are the excluded polar angles. The excluded solid angles for each nucleon are then given by
\begin{equation}
\Omega_{a}  =  2\pi (1-\cos \theta_A), \qquad \Omega_{b} =  2\pi (1-\cos \theta_B) \; ,
\label{eq10}
\end{equation}
and therefore the total allowed solid angle can be obtained from:               
\begin{equation}
\Omega_{pauli} = 4\pi-2(\Omega_a+\Omega_b-\bar \Omega) \; , 
\label{eq11} 
\end{equation}
\begin{figure}
\centering
\vspace*{0.2cm}
\hspace*{0.5cm}
\scalebox{0.6}{\includegraphics{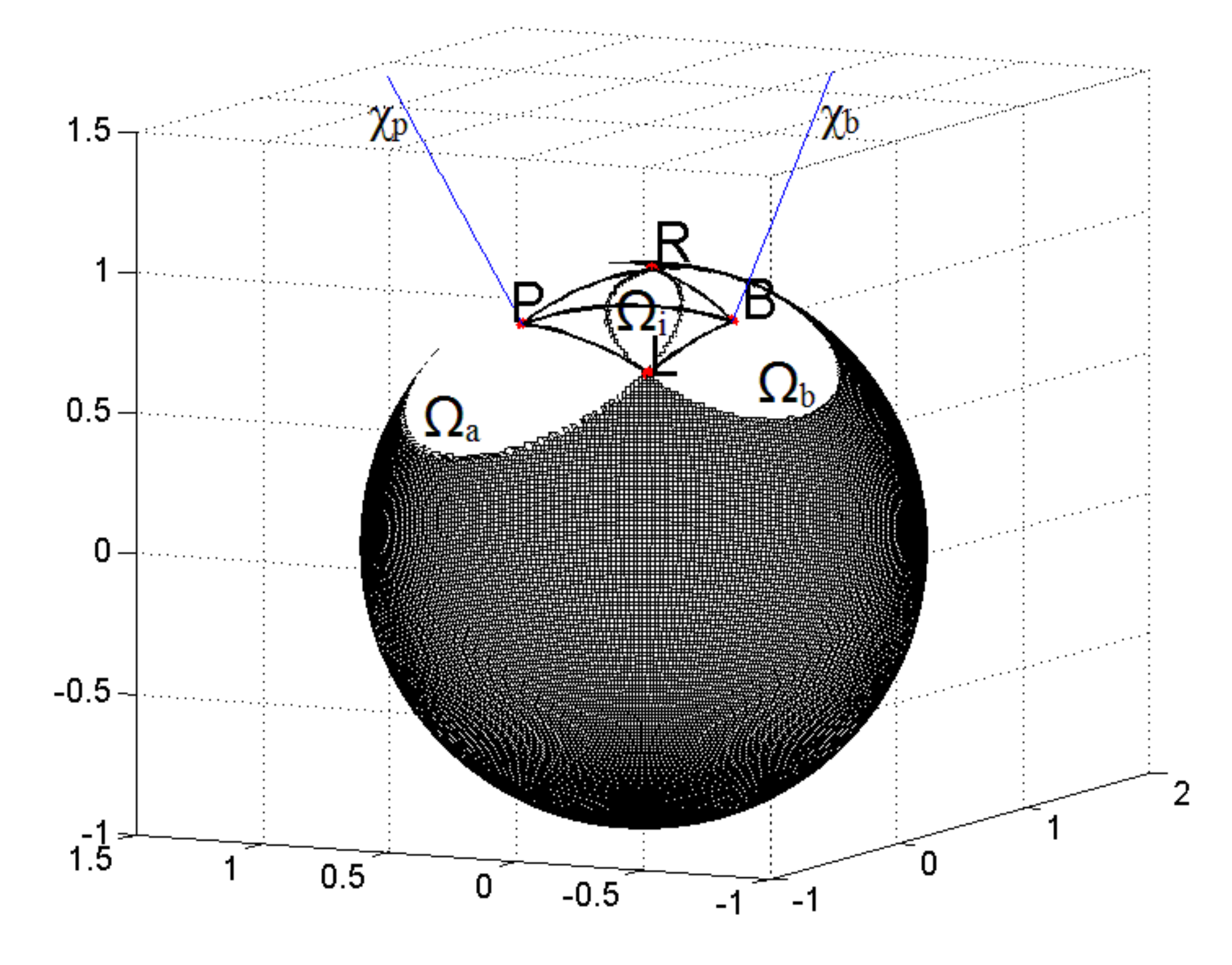}}
\caption{(Color online) Pauli blocking of two nucleons in three dimensions.}
\label{three}
\end{figure}
where $\bar \Omega$ represents the intersection  of the two conical sections $\Omega_a$ and $\Omega_b$. 
The full calculation has already been done in Ref.~\cite{Bert86}; however, in this paper we will use a slightly different 
approach to calculate $\bar \Omega$. Figure~\ref{three} shows how $\Omega_{a}$ and $\Omega_{b}$ are projected on 
the surface of a unit sphere. If $\Omega_i$ is the intersection of $\Omega_a$ and $\Omega_b$, it is obvious that
\begin{figure}
\centering            
\scalebox{0.42}{\includegraphics{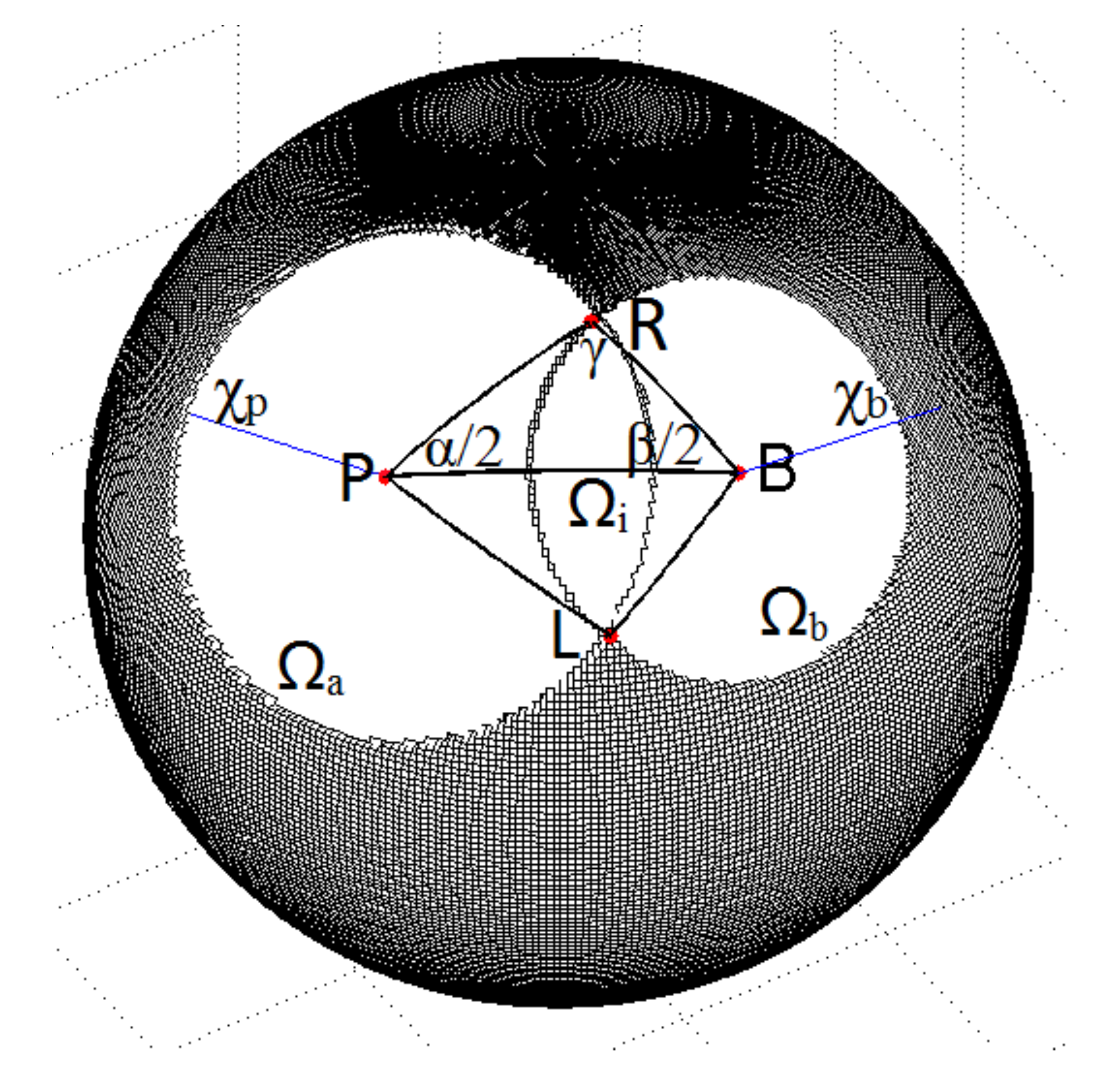}}
\caption{(Color online) A different view of Pauli blocking of two nucleons in three dimension}
\label{four}
\end{figure}
\begin{equation}
\Omega_{i} = \left\{ \begin{array}{ll} 0 & \mbox{if $\theta >\theta_{A} +\theta_{B}$} \; ; \\ \Omega_{b} & \mbox{if $\theta_{B} < \theta_{A}, \; \theta < |\theta_{B}-\theta_{A}|$} \; ; \\ \Omega_{a} & \mbox{ if $\theta_{A} < \theta_{B}, \theta < |\theta_{B}-\theta_{A}|$} \; .\end{array} \right.
\label{eq12}
\end{equation}
The case $ |\theta_{B} - \theta_{A}| < \theta <\theta_{A} +\theta_{B}$ is more complex than the other three cases and a more detailed study is needed. As shown in Fig.~\ref{four}, $P$ and $B$ are the centers of the two circular projections $\Omega_{a}$ and $\Omega_{b}$. The two circular contours intersect at $R$ and $L$. $\alpha/2$, $\beta/2$ and $\gamma$ are the internal angles of the spherical triangle PBR. 
The circular sectors of $\Omega_{a}$ and $\Omega_{b}$ have areas equal to ${\alpha}\Omega_{a}/({2\pi})$ and ${\beta}\Omega_{b}/({2\pi})$, respectively. The intersection area of $\Omega_{a}$ and $\Omega_{b}$ is given by
\begin{equation}
\Omega_{i} = \frac{\alpha}{2\pi}\Omega_{a}+\frac{\beta}{2\pi}\Omega_{b}-2\Delta_{PRB} \; . 
\label{eq13} 
\end{equation}
Here, $\Delta_{PRB}$ is the area of the spherical triangle PBR.
To obtain an expression for $\alpha /2$, first we define the center of the unit sphere, $O$, as the orgin of the system, and $\chi_{p}$ along the $z$-axis. Point B is at location $(1,\theta,\alpha/2)$, while point L has coordinates $(1,\theta_{A},0)$. We can then write:
\begin{equation}
{\bf {OB}} \cdot {\bf {OL}} = \cos \theta_{B} = \cos \theta_{A} \cos \theta + \sin \theta_{A} \sin \theta \cos(\alpha/2) \; ,
\label{eq14} 
\end{equation}
from which $\alpha/2$ can be readily obtained as
\begin{equation}
\alpha/2 = \arccos\left(\frac{\cos \theta_{B}-\cos \theta \cos \theta_{A}}{\sin \theta \sin \theta_{A}}\right) \; . 
\label{eq15} 
\end{equation}
In a similar fashion we find $\beta/2$ to be given by 
\begin{equation}
\beta/2 = \arccos\left(\frac{\cos \theta_{A}-\cos \theta \cos \theta_{B}}{\sin \theta \sin \theta_{B}}\right) \; . 
\label{eq16} 
\end{equation}
Applying the law of cosines of spherical trigonometry, 
\begin{equation}
\cos \gamma = -\cos(\alpha/2)\cos(\beta/2)+\sin(\alpha/2)\sin(\beta/2)\cos \theta \; , 
\label{eq17} 
\end{equation}
we obtain  
\begin{equation}
\gamma = \arccos(-\cos(\alpha/2)\cos(\beta/2)+\sin(\alpha/2)\sin(\beta/2)\cos \theta) \; . 
\label{eq18} 
\end{equation}
From Girard's theorem of spherical trigonometry, we have
\begin{equation}
\Delta_{PRB} = \alpha/2+\beta/2+\gamma-\pi \; . 
\label{eq19} 
\end{equation}
Inserting Eq.~(\ref{eq18}) and Eq.~(\ref{eq19}) into Eq.~(\ref{eq13}), the solid angle $\Omega_{i}$ is found to have the following value
\begin{equation}
\Omega_{i} = 2\{\pi-\cos\theta_{A}\cos^{-1}(\delta_{AB})-\cos\theta_{B}\cos^{-1}(\delta_{BA})-\cos^{-1}[\cos\theta\sqrt{(1-\delta_{AB}^{2})(1-\delta_{BA}^{2})}-\delta_{AB}\delta_{BA}]\} \; , 
\label{eq20} 
\end{equation}
where
\begin{equation}
\delta_{ij} = \frac{\cos\theta_{i}-\cos\theta\cos\theta_{j}}{\sin\theta\sin\theta_j} \; . 
\label{eq21} 
\end{equation}
Noticing that, while $\theta + \theta_{A} + \theta_{B} > \pi$, $\Omega_{a}$ and $\Omega_{b}$ have two intersections on the hemisphere, we have
\begin{equation}
\bar \Omega = \Omega_{i}(\theta,\theta_{A},\theta_{B})+\Omega_{i}(\pi-\theta,\theta_{A},\theta_{B}) \; . 
\label{eq22} 
\end{equation}

\begin{figure}
\centering            
\scalebox{0.38}{\includegraphics{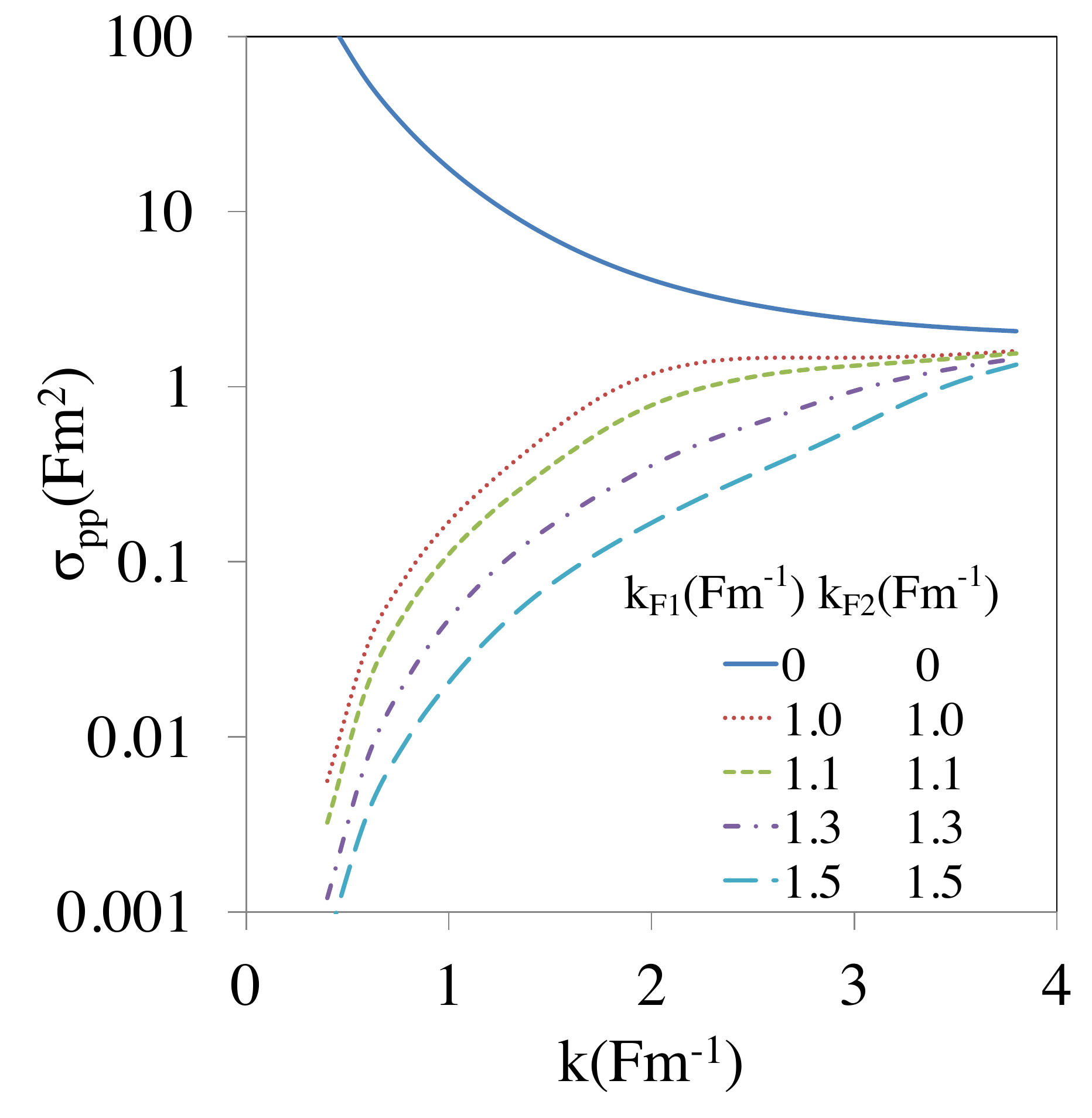}}\hspace{1.2cm}
\scalebox{0.38}{\includegraphics{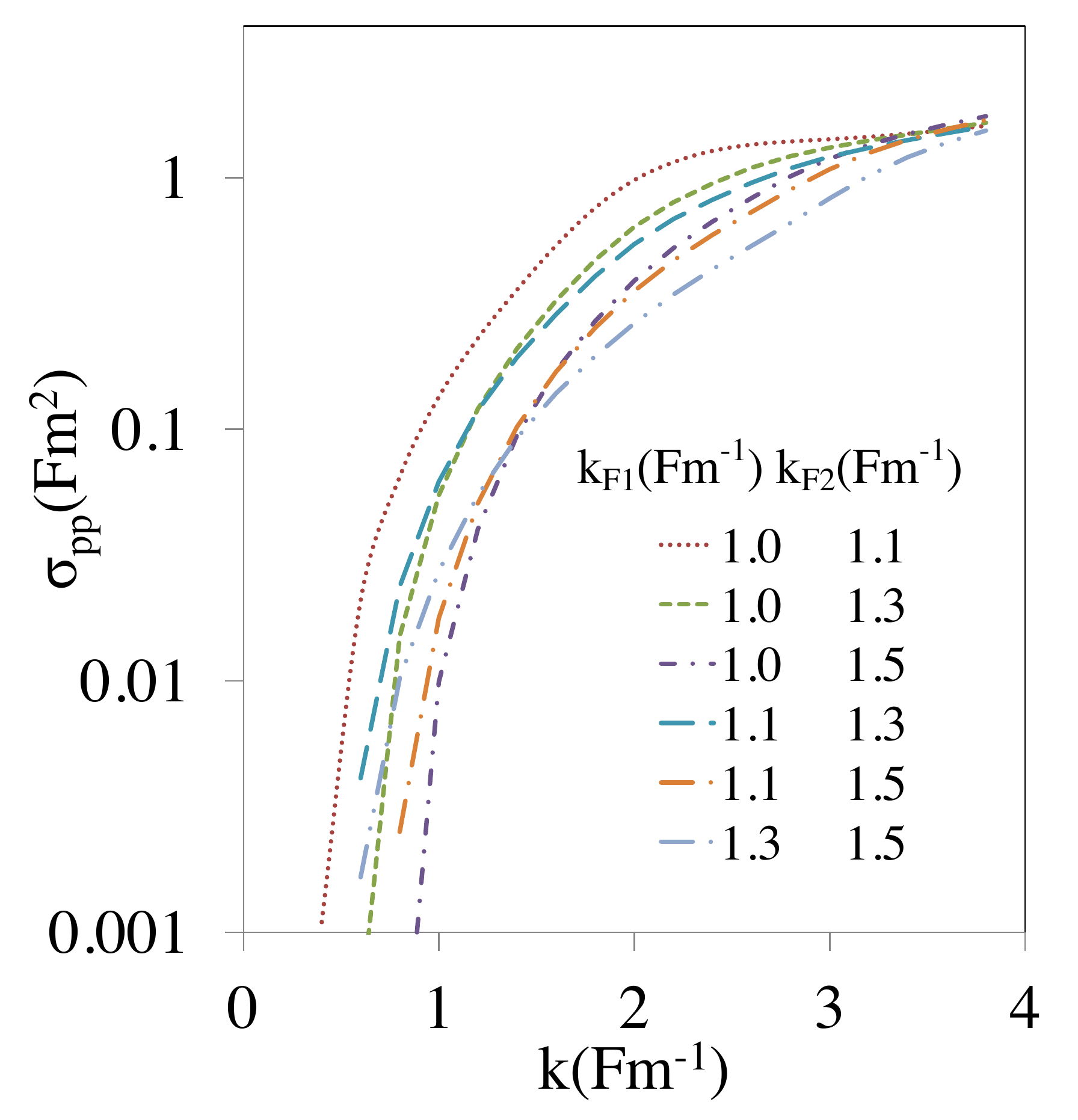}}
\caption{(Color online)                                                                                    
In-medium $pp$ cross section calculated as in Eq.~(\ref{sigav}) for a variety
of symmetric ($k_{F1}=k_{F2}$) and
asymmetric ($k_{F1}\ne k_{F2}$) situations. 
} 
\label{five}
\end{figure}

\begin{figure}[!t] 
\centering         
\scalebox{0.4}{\includegraphics{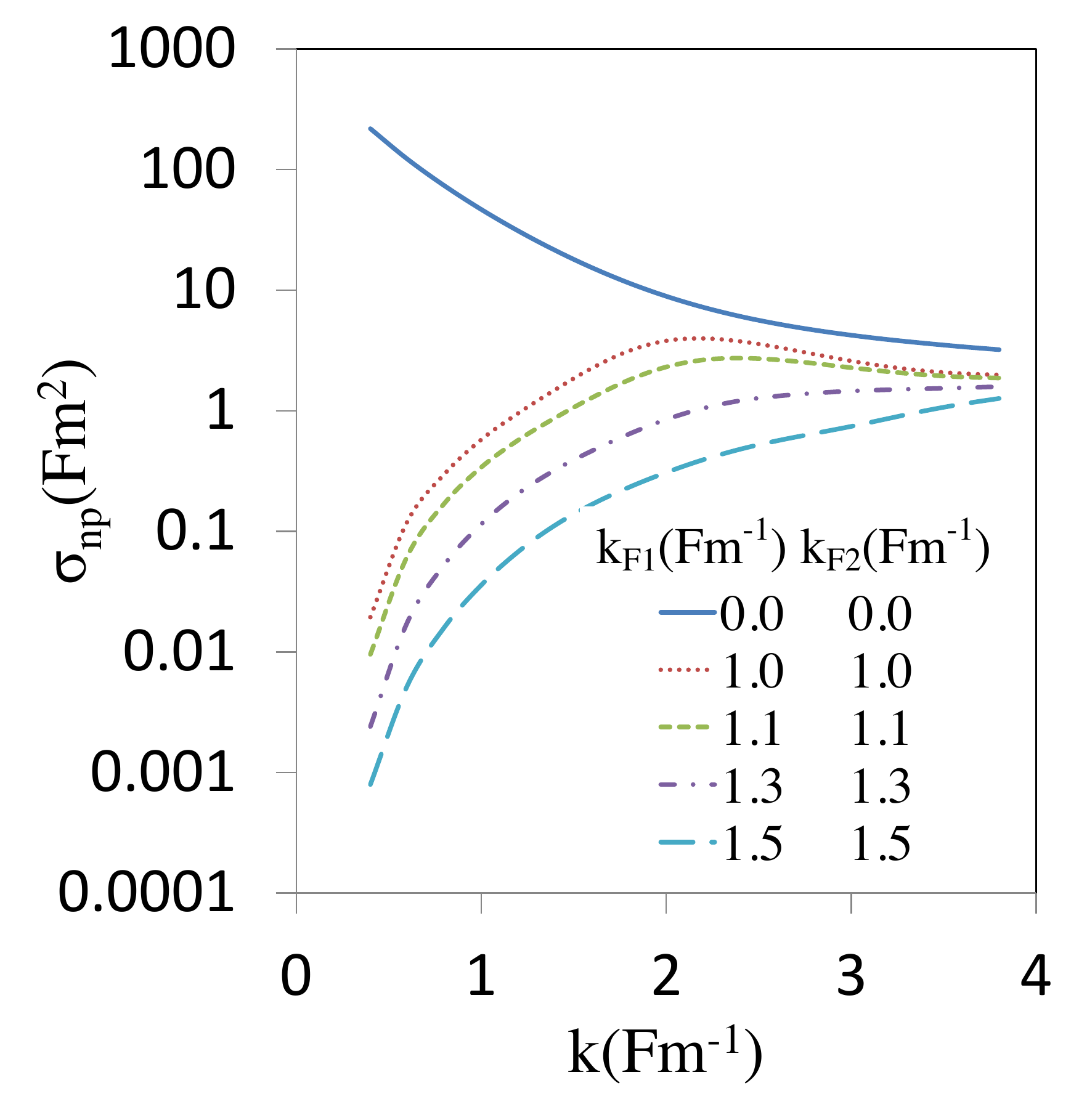}}\hspace{1.2cm}
\scalebox{0.4}{\includegraphics{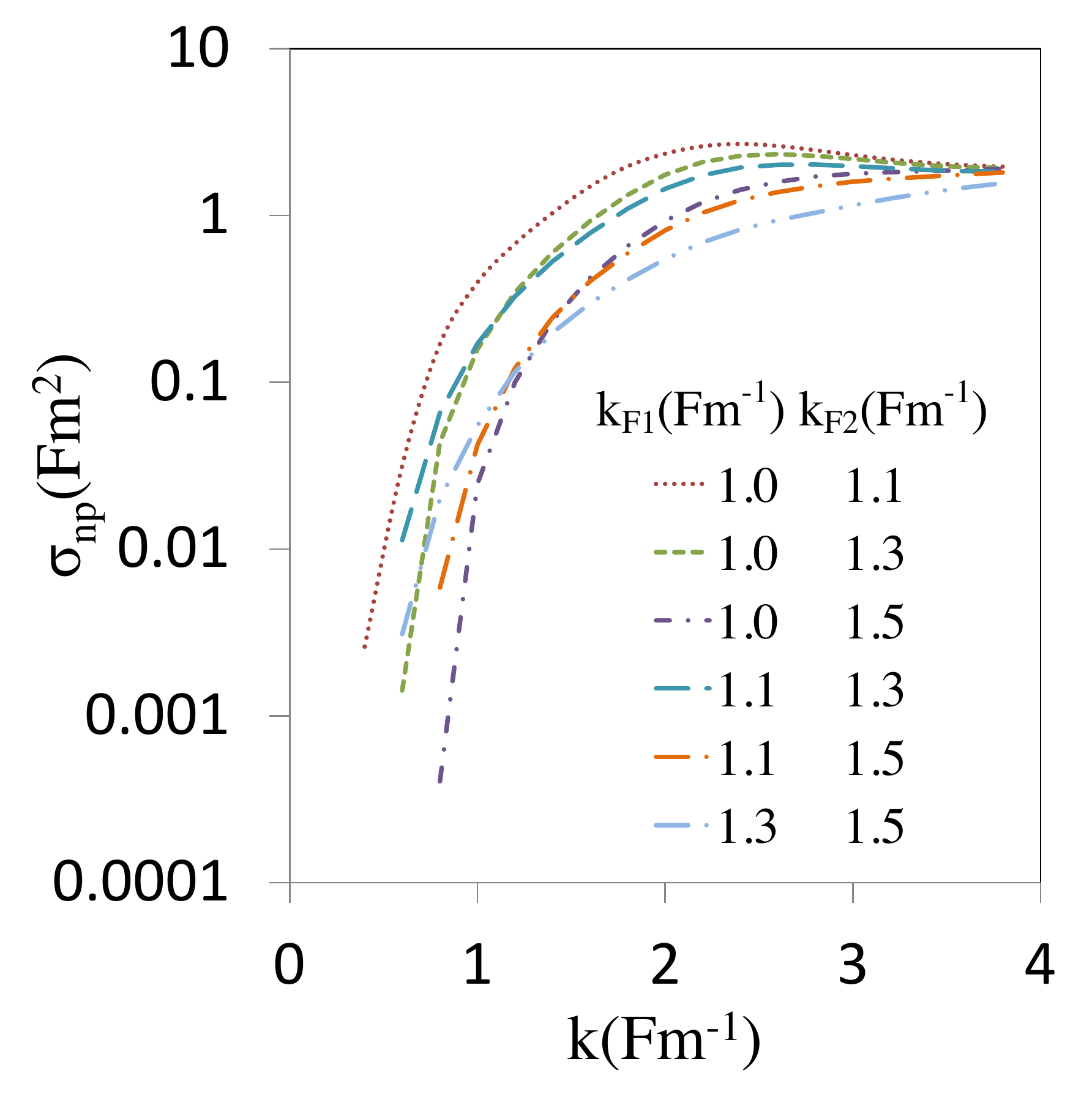}}
\caption{(Color online)                                                                                      
As in the previous figure, for $np$ scattering. 
} 
\label{six}
\end{figure}

\begin{figure}[!t] 
\centering         
\scalebox{0.4}{\includegraphics{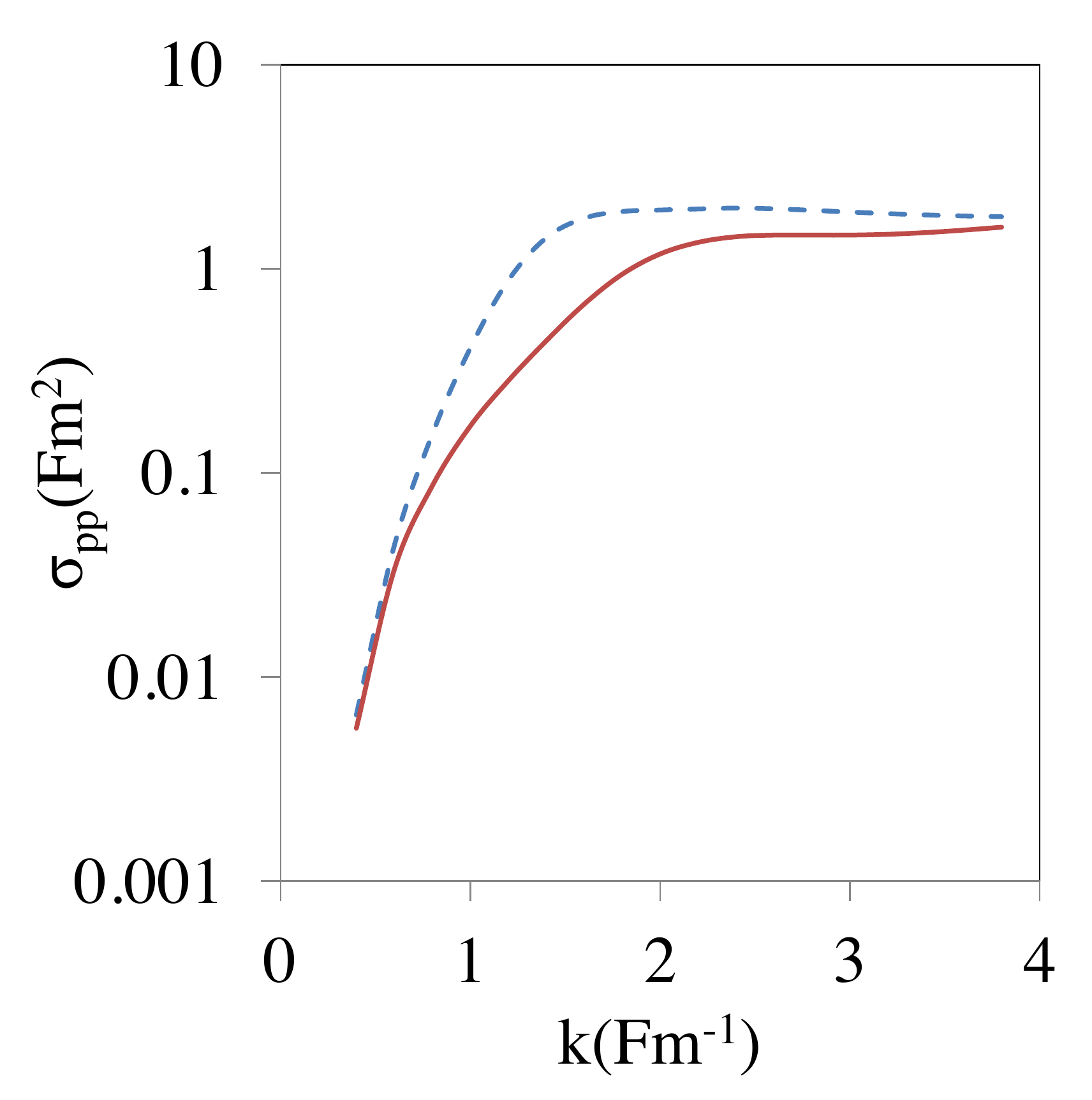}}\hspace{1.2cm}
\scalebox{0.4}{\includegraphics{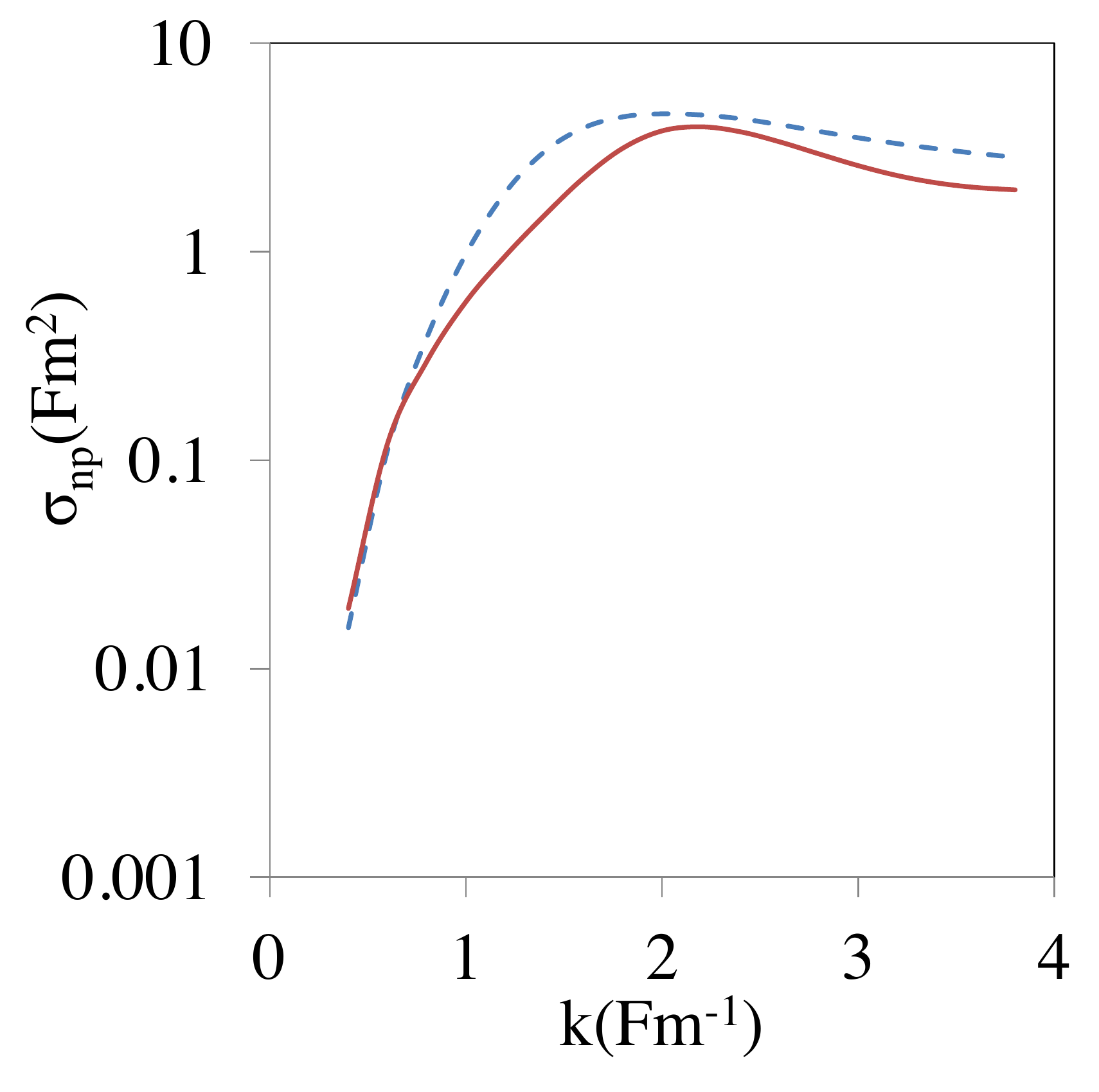}}
\caption{(Color online)                                                                                      
$pp$ (left) and $np$ (right) in-medium cross sections calculated from Eq.~(\ref{sigav}) 
with $k_{F1}=k_{F2}=1.0 \ {\rm fm}^{-1}$. 
Solid red: predictions as in Figs.~\ref{five}-\ref{six}; dashed blue: the input NN cross section
is evaluated in free space. See text for details. 
} 
\label{seven}
\end{figure}

\begin{figure}[!t] 
\centering         
\scalebox{0.38}{\includegraphics{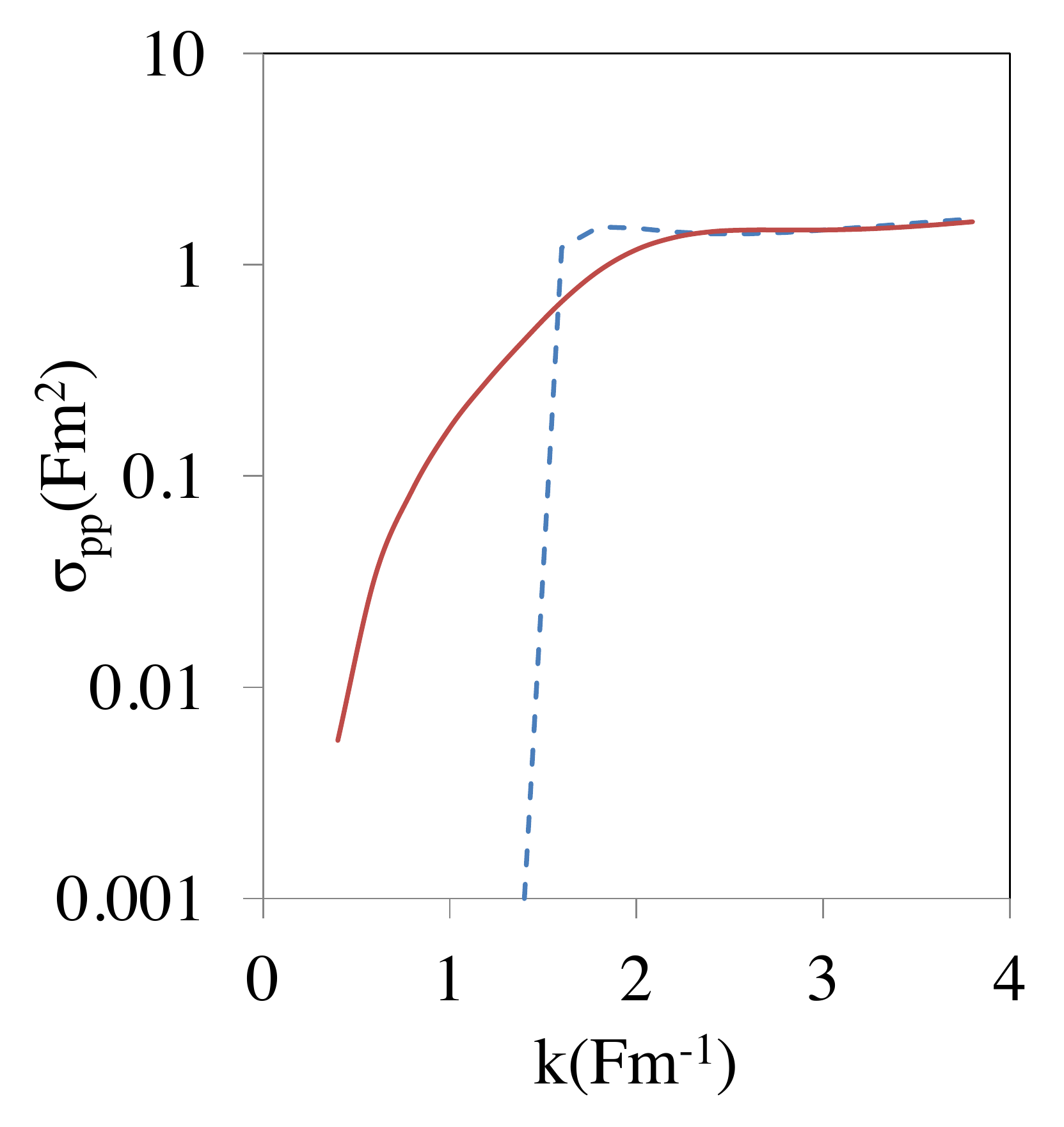}}\hspace{1.2cm}
\scalebox{0.38}{\includegraphics{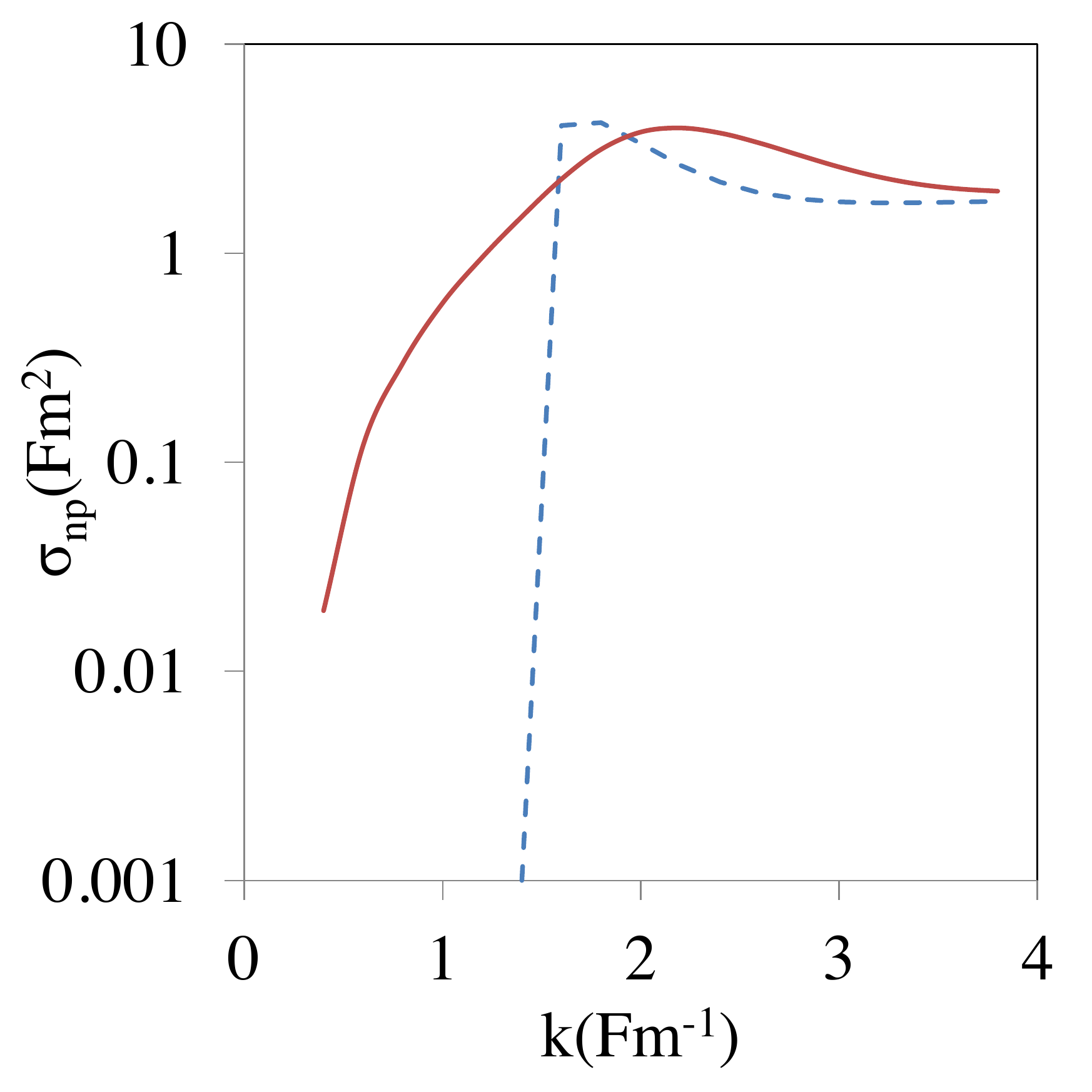}}
\caption{(Color online)                                                                                      
$pp$ (left) and $np$ (right) in-medium cross sections                          
with $k_{F1}=k_{F2}=1.0 \ {\rm fm}^{-1}$. 
Solid red: predictions as in Figs.~\ref{five}-\ref{six}; dashed blue: predictions obtained with  
Eq.~(\ref{sigmed}). 
} 
\label{eight}
\end{figure}

\section{Results} 
\subsection{Effective NN cross sections}

We begin by showing in Fig.~\ref{five} the average in-medium $pp$ cross section calculated as 
in Eq.~(\ref{sigav}). On the left, we display a variety of cases with equal Fermi momenta, whereas
asymmetric cases are shown on the right. Figure~\ref{six} contains the same information for the
$np$ cross section. After ``overcoming" complete Pauli blocking, the cross section generally
rises with increasing incident momentum.
In the $np$ case, we observe, at least at the lower densities, a tendency to reach a broad 
maximum. In all cases, the cross sections become nearly flat at the larger momenta and begin to approach
the free space predictions. 

Figures~\ref{five}-\ref{six} are more insightful when compared with Fig.~\ref{seven}. There, the $pp$ and $np$ cross sections shown 
by the dashed blue line 
are also calculated with Eq.~(\ref{sigav}), but the input NN cross sections in the integrand are evaluated in
free space. Thus, comparing the two curves on the left-hand-side (or on the right, for $np$)
shows the impact of the additional medium effects (besides those coming from the $\int_{Pauli} d \Omega$
factor in Eq.~(\ref{sigav})) originating from the G-matrix calculation and included in $\sigma_{NN}$ in the case of the 
solid curves. 
The impact is noticeable, with the microscopic medium effects further suppressing the cross section and      
shifting  the position of the peak.
We have chosen a particular case ($k_{F1}=k_{F2}=1.0 \ {\rm fm}^{-1}$) for the purpose of demonstration,
but the trend is similar for other densities.          

Figure~\ref{eight} is crucial for the point that we wish to make. There, for $pp$ and $np$ (on the left and right,
respectively), we compare the cross sections calculated from Eq.~(\ref{sigav}) to the corresponding ones evaluated
with Eq.~(\ref{sigmed}) instead. 
The predictions from Eq.~(\ref{sigmed}) have a sharper rise from zero and a more pronounced peak structure.
As is reasonable, 
differences are large at low momenta, where the scattering is most sensitive to the description
of Pauli blocking, particularly near the onset of the cross section.                                         
Again, we have taken a representative case, but this pattern is common to all densities. 
It will be interesting to explore the impact of such differences on reaction cross sections, our
next objective.

\section{Conclusions}                                                                  

Pauli blocking is perhaps the most important mechanism impacting the collision of two 
fermions in the medium. It is known to have a substantial effect on the scattering 
probability, that is, the in-medium cross section. 

In this paper, we predict in-medium effective NN cross sections suitable for applications to 
nucleus-nucleus scattering.                                             
The microscopic NN elastic cross sections, modified by all medium effects implied by the 
Dirac-Brueckner-Hartree-Fock theory of nuclear matter, 
are properly averaged so as to account for all possible directions of the relative momentum 
of two nucleons in the two colliding Fermi spheres. The more realistic description of the collision geometry amounts to an improved description of
Pauli blocking as compared to a previous approach \cite{SK}. 
We find the effective NN cross sections to be very sensitive to the description of the Pauli 
blocking geometry. 

Our future plans include the application of these cross sections in Glauber reaction calculations
with stable and unstable nuclei. 
In closing, we also note that in-medium cross sections are related to the mean-free path of a nucleon, a fundamental quantity in the 
description of a nucleon propagation in nuclear matter. 

\section*{Acknowledgments}
Support from the U.S. Department of Energy under Grant No. DE-FG02-03ER41270, DE-FG02-08ER41533 and DE-FG02-10ER41706 is 
acknowledged.                                                                

\end{document}